\documentclass[a4paper,aps,11pt,preprint]{revtex4}
\usepackage{graphicx,times,nicefrac,amsmath,amsfonts,amssymb,amsthm,epsf,bm,bbm,longtable,dcolumn}
\usepackage{revsymb,wasysym,mathrsfs,color}
\begin{document}

\title{Ionization-excitation of helium-like ions at Compton scattering}

\author{A. I. Mikhailov and A. V. Nefiodov}\thanks{E-mail: anef@thd.pnpi.spb.ru}
\affiliation{Theoretical Physics Division, Petersburg Nuclear Physics Institute, 188300 Gatchina, St. Petersburg, Russia}

\widetext
\begin{abstract}
Ionization of helium-like ions with simultaneous excitation of  the $ns$-states due to photon scattering is considered. The differential and total cross sections of the process are calculated to leading order of perturbation theory with respect to the interelectron interaction. The formulas obtained are applicable in the nonrelativistic energy range far beyond the ionization threshold.
\end{abstract}
\maketitle

\section{Introduction}

Scattering of photons from an atom accompanied by the transition of a bound electron to a continuous spectrum is usually called Compton scattering. The simplest target is a one-electron atom characterized by the ionization potential $I = m (\alpha Z)^2/2$ and the average momentum $\eta = m \alpha Z$ of a bound electron, where $\alpha$ is the fine-structure constant and $m$ is the electron mass ($ \hbar = 1 $, $ c = 1 $).  The atomic nucleus with  charge number $Z$ is assumed to be the source of an external field (Furry picture). In the nonrelativistic limit, the Coulomb parameter is $\alpha Z \ll 1$. The cross section for the Compton scattering of photons with energy $\omega_1$  in the range $I \ll \omega_1 \ll m$ was derived by Schnaidt \cite{1} and can be written as:
\begin{equation} \label{eq1}
\sigma^+_{1s} =\frac{8\sigma_{\mathrm{T}}}{\nu^2_1} \int\limits^{\varepsilon_1-1}_0
\frac{d\varepsilon}{1-e^{-2 \pi\xi}}\int\limits^{x_{\max}}_{x_{\min}} dx (1+\tau^2(x))F(x),
\end{equation}
where
\begin{gather*}
F(x)= \frac{x (3x+\Delta) e^{-\gamma(x)} }{[(x-\Delta)^2+4x]^3}, \\
\tau(x) = \frac12\left(\frac{\nu'_2}{\nu_1} + \frac{\nu_1}{\nu'_2}-\frac x{\nu_1\nu'_2}\right) , \\
\gamma(x)=2 \xi  \cot^{-1} \left(\frac{x+1-\varepsilon}{2\sqrt{\varepsilon}}\right) .
\end{gather*}
Here $\sigma_{\mathrm{T}}=8\pi r^2_e/3=6.65\times 10^{-25}$~cm$^2$ is the Thomson cross section for scattering from a free electron,  $r_e=\alpha/m$ is the classical electron radius,  $\omega'_2=\omega_1-E-I$ is the energy of scattered photon, $\varepsilon_1= \omega_1/I$ and $\varepsilon= E/I$ are  the energies of incident photon and ionized electron, respectively, calibrated in units $I$,  $\Delta=(\omega_1- \omega'_2)/I= \varepsilon+1$ is the dimensionless energy loss of inelastically scattered photon,  $x_{\min}=(\nu_1-\nu'_2)^2$, $x_{\max} =(\nu_1+\nu'_2)^2$,   $\nu_1=\omega_1/\eta = \alpha Z\varepsilon_1/2$,    $\xi=1/\sqrt{\varepsilon}$,  and $\nu'_2 =\omega'_2/\eta=\nu_1- \alpha Z\Delta/2$.  The range of the principal value of  $\cot^{-1}  z$ lies between $0$ and  $\pi$.

In view of the condition $\omega_1 \gg I$, the photon scattering is described within the framework of the $ {\bf  A}^2$-approximation \cite{2,3,4}, where ${\bf A}$ is the vector-potential of the photon field. In this approximation, formula \eqref {eq1} is deduced from the amplitude that corresponds to the contact (or sea-gull) Feynman diagram, with electrons, both bound and ejected into a continuous spectrum, described by the Coulomb wave functions. In the near-threshold energy range $\omega_1 \gtrsim I$, the contribution of  pole terms  also has to be taken into account  \cite{5}, but in this case the Compton cross section itself is sufficiently small in comparison with the photoabsorption cross section. The ionization cross sections for  photo-  and  Compton effects become comparable in magnitude at  photon energy $\omega_ {c} \simeq (5/7) \eta Z^{2/5}$.  In particular,  for $Z = 2$ we have $\omega_{c} \simeq \eta \simeq 7$~keV. If  $\omega_1 \gg \eta$, the ionization occurs mainly due to the Compton scattering.

\begin{figure}[tp]
\begin{minipage}[t]{0.49\textwidth}
\centering\includegraphics[width=\textwidth,angle=0,clip]{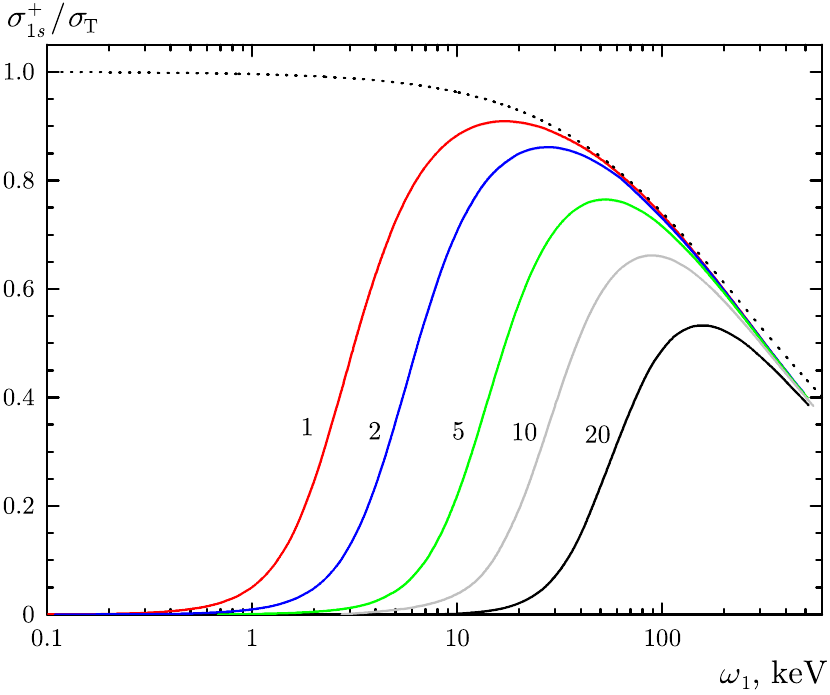}
\end{minipage}
\caption{\label{fig1} Cross sections  for the Compton scattering from a bound $K$-electron are calculated by formula \eqref {eq1} (solid curves). The numbers indicate the charge number $Z$. The dotted curve corresponds to scattering from a free electron.}
\end{figure}

The dependence of  $\sigma^+_{1s}$ on the energy $\omega_1$ of  incident photons for various one-electron atoms is presented in Fig.~\ref{fig1}. The cross section of the Compton effect on a free electron calculated by the Klein-Nishina-Tamm formula \cite{6} is also  shown there for comparison.  At $\omega_1 \ll \eta$, the cross section for the Compton scattering on bound electron is small, since the process occurs in the region kinematically 
inaccessible for scattering on free electron. The photon energy losses  in free kinematics determined by the conservation laws of energy and momentum are uniquely linked with the scattering angle $\vartheta$ by the relation (in units of $I$) \cite{6}:
\begin{equation}
\delta= 2(1 -\cos \vartheta)\omega^2_1/\eta^2 . 
\end{equation}
Even the maximum values of these losses $\delta_{\max} \simeq 4 \omega^2_1 / \eta^2 \ll 1$, which are achieved by backward scattering ($\vartheta \simeq \pi $), are too small in comparison with the photon energy losses $\Delta = \varepsilon + 1 \geqslant 1$  in the Compton scattering from atom [see formula \eqref {eq1}]. If the interaction of an electron with a nucleus is taken into account, only the energy-conservation law takes place. The atomic nucleus participates in the process, absorbing any recoil momentum by virtue of the enormous mass.

\noindent

At  $\omega_1 \sim \eta $, the ionization of  bound electron occurs most efficiently. In this case, the energy loss of  photon scattered from an atom is close to the energy loss $\delta \sim 2 (1 - \cos \vartheta)$ in free kinematics in a wide range of scattering angles, except for the scattering at small angles 
$\vartheta  \sim  0$.

At  $\eta \ll \omega_1 \ll m $, the photon mainly loses the energy $\Delta \sim 2 \omega ^ 2_1 / \eta^2 \gg 1$  \cite{4}. Accordingly, since the energy of ionized electron $\varepsilon \simeq \Delta$, the Coulomb parameter is $\xi = 1 / \sqrt{\varepsilon} \sim \eta / \sqrt{2} \omega_1 \ll 1$ and the wave function of the electron can be approximated by a plane wave (Born approximation). This gives  instead of Eq.~\eqref{eq1} the cross section  $\sigma^+_{1s} = \sigma_{\mathrm {T}} (1-2 \omega_1 / m) $, which does not depend on $Z$.

The relativistic expressions for differential cross sections of  the Compton scattering from a $K$-electron were studied in a number of papers (see, for example, works \cite {7,8,9,10}).  Formula \eqref {eq1} remains valid in the relativistic energy range $\omega_1 \sim m$, if the ionized electron is nonrelativistic \cite{11}. We also note that the cross section for two-electron atom (helium-like ion) to leading order of perturbation theory with respect to the interelectron interaction is given by $\sigma^+ = 2 \sigma^+_{1s}$, taking into account the number of electrons in the atom.

When studying the problem of scattering by atomic targets with a few electrons, one distinguishes such processes  that are entirely due to the interelectron interaction. The cross sections turn out to be extremely sensitive to the correct description of  the electron-electron correlations. The theoretical predictions made within the framework of different methods sometimes diverge from each other even by an order of magnitude.

\begin{figure}[t]
\begin{minipage}[t]{0.55\textwidth}
\centering\includegraphics[width=\textwidth,angle=0,clip]{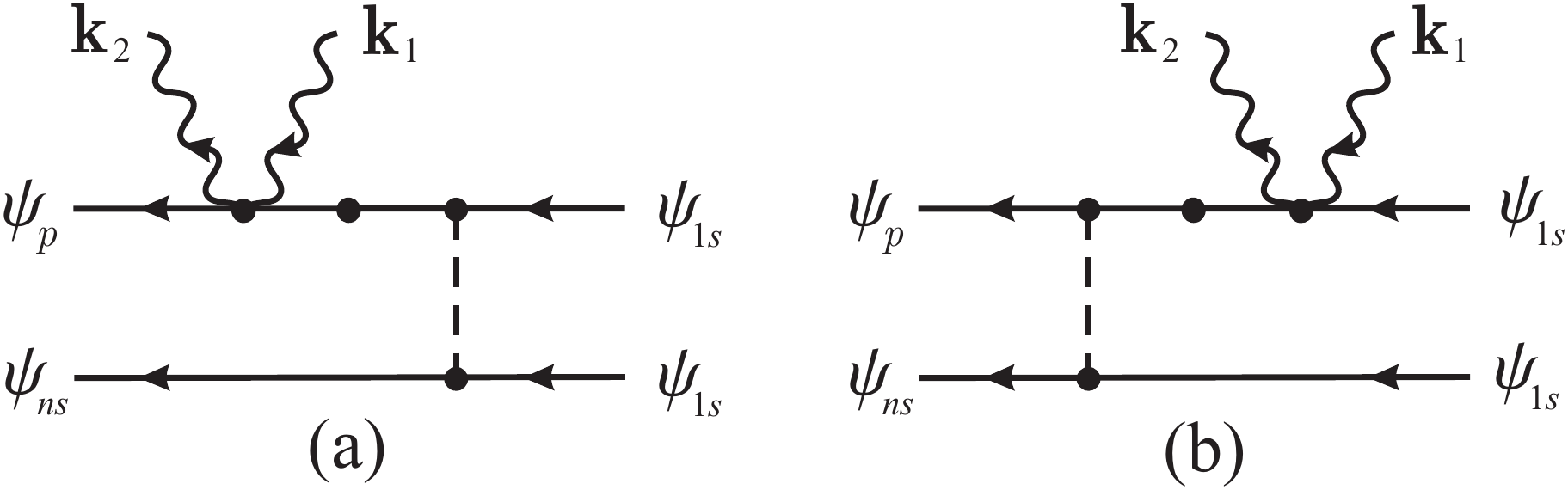}
\end{minipage}
\caption{\label{fig2} Contact Feynman diagrams for the ionization-excitation of an atom at photon scattering. The wavy lines represent the incident and scattered photons with momenta ${\bf  k}_1$ and ${\bf  k}_2$, respectively. The dashed line represents the electron-electron Coulomb interaction. The electron propagator with a point corresponds to the Coulomb Green's function. }
\end{figure}

In this paper, we shall consider the ionization of  two-electron atomic target with the simultaneous excitation of  the residual ion (ionization-excitation) into the  $ns$-state ($n \geqslant 2$) due to the scattering of photons with energy in the range $I \ll \omega_1 \ll m$. The incident photon interacts with a single electron only, so that the simultaneous transition of two bound electrons is possible merely, if the interelectron interaction is taken into account. We describe this interaction to the first order of  nonrelativistic perturbation theory with respect to the parameter $1/Z  \ll 1$, using the Coulomb wave functions and the Coulomb Green's function as the zeroth-order approximation. The Feynman diagrams for the process under consideration within the framework of the ${\bf A}^2$-approximation are depicted in Fig.~\ref{fig2}, where  graph \ref{fig2}a takes into account the electron-electron interaction in the initial state of  atom, while  graph \ref{fig2}b does it in the final state. To the diagrams in  Fig.~\ref{fig2}, one needs to add two more exchange diagrams, which are obtained from \ref{fig2}a  and   \ref{fig2}b  by permutation of the final states.

The ionization-excitation process at the Compton scattering was studied earlier in papers \cite {12,13,14} in the asymptotic nonrelativistic energy range $\eta \ll \omega_1 \ll m$. In this energy range, the problem allows for significant simplifications. The dominant contribution to the cross section of the process arises from the diagram in  Fig.~\ref {fig2}a only, and the Born approximation can be used to describe the wave function of emitted electron. The ratio of the ionization-excitation cross section $\sigma^{+ *}_{nl}$ into the $nl$-state to the ordinary ionization cross section is of  experimental interest:
\begin{equation}\label{eq3}
R_{nl}=\frac{\sigma^{+*}_{nl}}{\sigma^{+}} =  \frac{Q_{nl}}{Z^{2}}. 
\end{equation}
Here the dimensionless function $Q_{nl}$ does not depend on $\omega_1$ or $Z$, while  $\sigma^+ = 2\sigma_{\mathrm{T}}(1-2\omega_1/m)$.  
In particular, for the states with $n=2$,  $Q_{2s}= 0.0592$  and $Q_{2p}= 0.0043$ were calculated  \cite{13,14}. Universal scaling  \eqref{eq3} is obtained within the framework of nonrelativistic perturbation theory to leading order with respect to the parameter  $1/Z$ in the energy range  $\eta \ll \omega_1 \ll m$ only. As we shall see later, if the energy range is extended up to $I \ll \omega_1 \ll m$, which includes $\omega_1 \sim \eta$, the universality is violated and functions $Q_{nl}$ become explicitly dependent on $Z$ and $\omega_1$, herewith $\sigma^+ = 2 \sigma^+_{1s}$, where 
$\sigma^+_{1s}$ is given by Schnaidt's formula \eqref{eq1}.

In the literature, there are only two papers \cite {15,16} of  the same group of authors, where the ionization-excitation cross section for the helium atom  in the energy range $\omega_1  \sim  \eta$  was calculated. However, the approximations used in the calculations  \cite{15,16}, in our opinion, are not justified. Firstly, the so-called ``impulse approximation'' was used, in which the cross section is represented as a product of the Klein-Nishina-Tamm cross section and the atomic form factor. Secondly, the electron-electron interaction in the final state of  atom, which, in the energy range 
$\omega_1 \sim \eta$, contributes the same order of magnitude as the interaction in the initial state, was not taken into account.

In paper \cite{17}, we used nonrelativistic perturbation theory in order to describe the ionization-excitation of helium-like targets at scattering of fast electrons (with energy much higher than the binding energy $I$). Since, in this case, the dominant contribution to the ionization cross section is caused by small energy losses (of the order of $I$), the interaction of  fast particle with an atom can be described by the operator \cite {17}
\begin{equation}\label{eq4}
U({\bf r}) = \frac{4\pi\alpha}{q^2} e^{i{\bf q}\cdot {\bf r}} ,
\end{equation}
where ${\bf q}$ is the momentum transferred by projectile electron. It is interesting that the radial dependence of the electron-photon interaction operator $U_\gamma ({\bf  r})$ within the framework of the ${\bf A}^2$-approximation has the same form; only the pre-exponential factor changes:
\begin{equation}\label{eq5}
U_\gamma({\bf r})=N_\gamma e^{i{\bf q}\cdot {\bf r}}, \quad
N_\gamma=2 \pi\frac{ \alpha}{m} \frac{\bm{\mathrm{e}}^*_2 \cdot \bm{\mathrm{e}}_1}{\sqrt{\omega_1 \omega_2 }}   .
\end{equation}
Here  ${\bf q}={\bf k}_1-{\bf k}_2$ is the momentum transferred to an atom by an incident photon, $\omega_1=|{\bf k}_1|$ and ${\bf e}_1$ ($\omega_2=|{\bf k}_2|$ and ${\bf e}^*_2$) are the energy and polarization vector of  incident (scattered) photon, respectively. Using the analogy between Eqs.~\eqref{eq4} and \eqref{eq5}, it is easy to reconstruct the amplitude of the process under investigation from the results of  work \cite{17}.

\section{Amplitude for ionization-excitation of  atom at Compton scattering}

Let us denote by $p$ and $E = p^2/2m$ the asymptotic momentum and energy of  ionized electron, while by $\eta_n = \eta /n$ and $E_{ns} = - \eta^2_n/2m$ the average momentum and energy of  excited electron, respectively $(n \geqslant 2)$. Following \cite{17}, we represent the required amplitude as the sum of four matrix elements corresponding to contributions from four Feynman diagrams, the first two of which are shown in 
Fig.~\ref{fig2}:
\begin{equation}\label{eq6}
{\cal A} = \sqrt{2} \left({\cal A}_a + {\cal A}_b + {\cal A}_c +  {\cal A}_d \right) . 
\end{equation}
Here 
\begin{gather}
{\cal A}_a = \langle\psi_p\psi_{ns}|U_\gamma G(E_a)V|\psi_{1s}\psi_{1s}\rangle, \\
{\cal A}_b= \langle\psi_p\psi_{ns}| VG(E_b) U_\gamma |\psi_{1s}\psi_{1s}\rangle, \\
{\cal A}_c = \langle\psi_{ns}\psi_p| U_\gamma G(E_c) V |\psi_{1s}\psi_{1s}\rangle, \\
{\cal A}_d = \langle \psi_{ns}\psi_p| VG(E_b)U_\gamma |\psi_{1s}\psi_{1s}\rangle ,   
\end{gather}
where $G(E)=(E-H)^{-1}$ is the Coulomb Green's function for electron with the energy $E$. The electron-electron interaction is described by the two-particle operator  $V$,  while $U_\gamma$ and $G(E)$ are the single-particle operators. The energy of electrons in the intermediate states described by the Green's functions are determined by the energy-conservation law:
\begin{gather*}
E_a=2E_{1s}-E_{ns}=-I(2-n^{-2}), \\
E_b = E+E_{ns}-E_{1s}=I(\varepsilon+1-n^{-2}),  \\
E_c = 2E_{1s}-E = -I(\varepsilon+2),
\end{gather*}
where $\varepsilon=E/I=p^2/\eta^2$ is the dimensionless energy of  ionized electron. 

It should be noted that the amplitudes ${\cal A}_a$ and ${\cal A}_c$ account for the interaction between atomic electrons in the initial state, while the amplitudes ${\cal A}_b$ and ${\cal A}_d$  do it in the final state. The technique for calculating the amplitudes is described in detail in \cite {17}. Here we give only their final expressions. The direct amplitudes are represented as derivatives of single integrals:
\begin{gather}
{\cal A}_a=  {\cal N} \hat\Gamma_{\mu\lambda} \int\limits^1_0 \frac{dx}\Lambda e(x) \Phi(\Lambda,\lambda)
 {}_{\big|{\lambda\to 0 \hfill \atop \mu=\eta+\eta_n}} ,  \label{eq7}\\
 {\cal A}_b=  {\cal N} \hat\Gamma_{\mu\lambda} \int\limits^1_0 \frac{dx}{\Lambda_1} e_1(x)\Phi_1(\Lambda_1,\mu)
  {}_{\big|{\lambda=\eta \hfill \atop \mu=\eta+\eta_n}}.  \label{eq8}
\end{gather}
The differential operator  $\hat\Gamma_{\mu\lambda}$ acts on the parameters $\mu$ and $\lambda$, on which the integrand functions depend:
\begin{gather*}
\hat\Gamma_{\mu\lambda} =D_\mu\frac{\partial^2}{\partial\mu\partial\lambda}\frac1{\mu^2}, \\
D_\mu=\sum^{n-1}_{l=0} \frac{(n-1)!(2\eta_n)^l}{(n-l-1)!l!(l+1)!}
\frac{\partial^l}{\partial\mu^l},  \\
\Lambda=\sqrt{p_a^2(1-x)+(\mu+\eta)^2x},   \\
 \Lambda_1=\sqrt{(q^2x-p^2_b)(1-x)+\lambda^2x-i0},   \\
e(x)=x^{-\zeta} \bigg(\frac{\Lambda+p_a}{\mu+\eta+p_a}\bigg)^{2\zeta}, \\ 
 e_1(x)=x^{-i\beta}\bigg(\frac{(qx)^2+(\Lambda_1-ip_b)^2}{q^2+(\lambda-ip_b)^2}\bigg)^{i\beta},   \\
\Phi(\Lambda,\lambda)= \frac{[({\bf q}-{\bf p})^2+(\Lambda+\lambda)^2]^{i\xi-1}}{[q^2+(\Lambda+\lambda-ip)^2]^{i\xi}},  \\
\Phi_1(\Lambda_1,\mu) = \frac{[({\bf q} x-{\bf p})^2+(\Lambda_1+\mu)^2]^{i\xi-1}}{[(q x)^2
  + (\Lambda_1+\mu-ip)^2]^{i\xi}}, \\
 p_a =\sqrt{2m|E_a|}=\frac{\eta}{\zeta},   \quad  p_b=\sqrt{2mE_b}=\frac{\eta}{\beta},   \\
 \zeta=\frac{1}{\sqrt{2-n^{-2}}},  \quad    \beta=\frac{1}{\sqrt{\varepsilon+1-n^{-2}}}, \quad \xi\ = \frac\eta{p} =\frac{1}{\sqrt{\varepsilon}}.
\end{gather*}
In equations \eqref{eq7} and  \eqref{eq8}, after taking derivatives one should tend  $\lambda$ to $0$ and $\eta$, respectively,
and  set $\mu$  equal to $\eta + \eta_n$. 

The exchange amplitudes turn out to be more complicated and are represented in terms of  derivatives of twofold integrals:
\begin{gather}
{\cal A}_c=\frac{{\cal N}}{2} \hat\Gamma_{\mu\lambda\tau} \int\limits^1_0 \frac{dx}{\Lambda_2}e_2(x)
  \int\limits^1_0 \frac{dy}{L_2} W(L_2){}_{\big|{{\mu=\eta_n \hfill \atop \lambda=\tau=\eta}}} ,  \label{eq9}\\
{\cal A}_d= \frac{{\cal N}}{2}  \hat\Gamma_{\mu\lambda\tau} \int\limits^1_0 \frac{dx}{\Lambda_1} e_1(x)
  \int\limits^1_0 \frac{dy}{L_1}W(L_1) {}_{\big|{{\mu=\eta_n \hfill \atop \lambda=\tau=\eta}}} .  \label{eq10}
\end{gather}
Here
\begin{gather*}
\hat\Gamma_{\mu\lambda\tau}=D_\mu \frac{\partial^3}{\partial\mu\partial\lambda\partial\tau}, \\
\Lambda_2=\sqrt{(p_c^2+q^2x)(1-x)+\mu^2x}, \\
e_2(x)=x^{-\gamma} \bigg(\frac{(qx)^2+(\Lambda_2+p_c)^2}{q^2+(\mu+p_c)^2}\bigg)^\gamma, \\
W(L)\ =\ \frac{[({\bf q} xy-{\bf p})^2+(L+\tau)^2]^{i\xi-1}}{[(qxy)^2+(L+\tau-ip)^2]^{i\xi}}, \\
L_1=\sqrt{(qx)^2 y(1-y)+(\Lambda_1+\mu)^2y}, \\
L_2=\sqrt{(qx)^2 y(1-y)+ (\Lambda_2+\lambda)^2y},  \\
 p_c=\sqrt{2m|E_c|}=\frac\eta\gamma,  \quad \gamma=\frac1{\sqrt{\varepsilon+2}}.
\end{gather*}

In formulas \eqref{eq9} and \eqref{eq10}, the derivatives are evaluated at points $\mu=\eta_n$, $\lambda = \eta$, and 
$\tau=\eta$.  Amplitudes \eqref{eq7}--\eqref{eq10} contain the common factor
\begin{gather}\label{eq11}
{\cal N} =(4\pi)^3 \alpha^2 N_p N_n N_1^2 \frac{{\bf e}_2^*\cdot {\bf e}_1}{\sqrt{4\omega_1\omega_2}}, \\
N^2_p = \frac{2\pi\xi}{1-e^{-2\pi\xi}}, \quad N^2_n= \frac{\eta_n^3}\pi .
\end{gather}

\section{Differential and total cross sections}

The differential cross section of the process averaged over the polarizations of  incident photons and summed over the polarizations of  scattered photons is connected to amplitude \eqref{eq6} by the relation
\begin{equation}\label{eq13}
d\sigma^{+*}_{ns} =  2\pi \overline{\vphantom{| \cal{A}|^)} |\cal{A}|\,}\!{}^2 \delta(\omega_2+E+E_{0}-\omega_1)
 \frac{d{\bf k}_2}{(2\pi)^3} \frac{d{\bm{\mathrm{p}}}}{(2\pi)^3},
\end{equation}
where
\begin{equation}\label{eq14}
\overline{\vphantom{| \cal{A}|^)} |\cal{A}|\,}\!{}^2 = \frac{1}{2}\sum_{\mathrm{polar}} |{\cal{A}}|^2 , 
\end{equation}
and $E_0=E_{ns}-2E_{1s}=I(2-n^{-2})$ is the threshold energy of the process. Taking into account that the amplitude  ${\cal{A}}$ depends on  angles via $q$, $({\bf p}\cdot{\bf q})$, and $({\bf e}^*_2\cdot{\bf e}_1)$, we represent the phase volumes in the form:
\begin{equation}
d{\bf k}_2=2\pi \omega^2_2d\omega_2dt_{12},  \quad   d{\bf p}=2\pi mpdEdt . 
\end{equation}
Here $t_{12}=\cos\theta_{12}$, $t=\cos\theta$, where $\theta_{12}$ is the angle between ${\bf k}_1$ and ${\bf k}_2$, $\theta$ is the angle between ${\bf p}$ and ${\bf q}={\bf k}_1-{\bf k}_2$. Eliminating the $\delta$-function in  \eqref{eq13} by integrating with respect to  variable  $\omega_2$  and replacing $dt_{12}$ by $qdq/\omega_1\omega_2$, we obtain
\begin{equation}\label{eq15}
d\sigma^{+*}_{ns} = \frac{mp}{(2\pi)^3} \overline{\vphantom{| \cal{A}|^)} |\cal{A}|\,}\!{}^2 
\frac{\omega_2}{\omega_1} dE q dq dt ,
\end{equation}
where  $\omega_2=\omega_1-E-E_0$.

It is further convenient to express amplitude \eqref{eq6} in terms of  the dimensionless quantity ${\cal{M}}$:
\begin{equation}
{\cal{A}}=\sqrt{2}\sum_k  {\cal{A}}_k= \sqrt{2} \eta^{-7}  {\cal N}  {\cal{M}} ,
\end{equation}
where factor ${\cal N}$  is determined by formula \eqref{eq11}.  The momenta and energies involved in ${\cal{M}}$ are expressed in the units of 
$\eta=m\alpha Z$ and $I=m(\alpha Z)^2/2$, respectively, and the energy-conservation  law  in these units takes the form $\varepsilon_1-\varepsilon_2=\varepsilon+\varepsilon_0$, where $\varepsilon_{1,2}=\omega_{1,2}/I$, $\varepsilon_0=2-n^{-2}$.  As a result, for given energy $\omega_1$  of an incident photon, quantity ${\cal{M}}$ depends on three dimensionless variables:  $\varepsilon=E/I=\xi^{-2}$, $\varkappa=q/\eta$, and $t$.

\begin{figure}[tb]
\begin{minipage}[t]{0.49\textwidth}
\centering\includegraphics[width=0.9\textwidth,angle=0,clip]{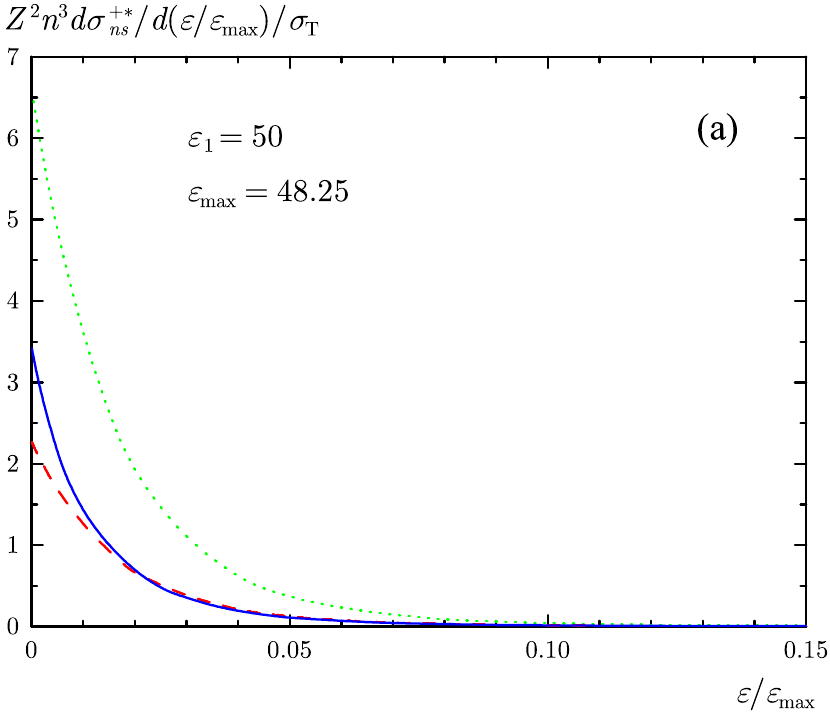}
\end{minipage}
\begin{minipage}[t]{0.49\textwidth}
\centering\includegraphics[width=0.9\textwidth,angle=0,clip]{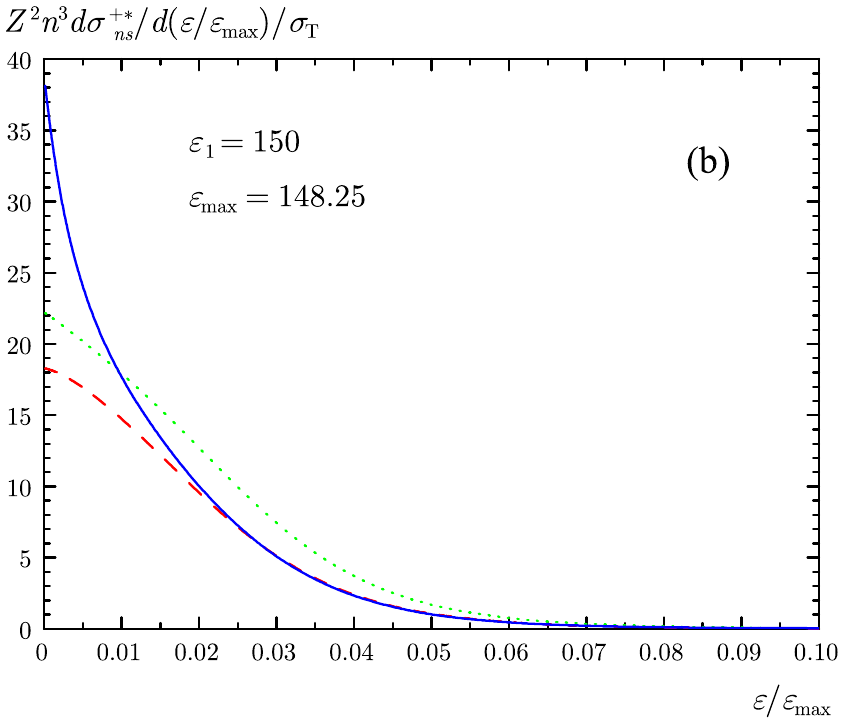}
\end{minipage}
\begin{minipage}[t]{0.49\textwidth}
\centering\includegraphics[width=0.9\textwidth,angle=0,clip]{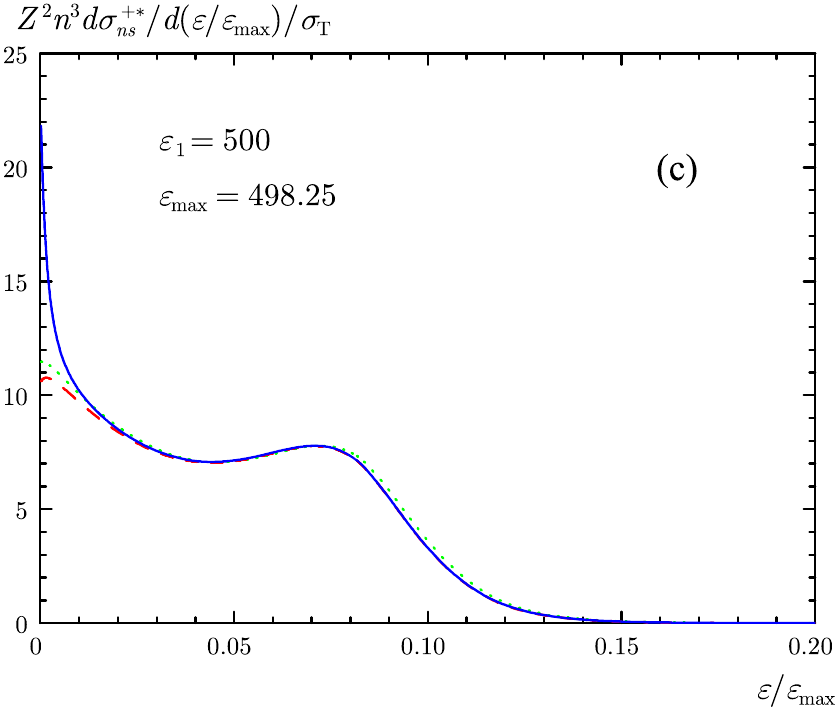}
\end{minipage}
\caption{\label{fig3} Energy distributions of  emitted electrons  for   $n=2$  and  $Z=2$: the contribution of the diagram in Fig.~\ref{fig2}a (dotted curve), the total contribution of the diagrams in Fig.~\ref{fig2}a  and  \ref{fig2}b (dashed curve), and the total contribution of all diagrams (solid curve).}
\end{figure}

Performing the summation over photon polarizations, we obtain
\begin{gather*}
\frac12 \sum_{\mathrm{polar}} |{\bf e}_2^*\cdot{\bf e}_1|^2= \frac12(1+t^2_{12}) , \\
t_{12}=\frac12 \left(\frac{\omega_2}{\omega_1} +\frac{\omega_1}{\omega_2}
  - \frac{q^2}{\omega_1\omega_2}\right) = \frac12 \left(\frac{\nu_2}{\nu_1} +\frac{\nu_1}{\nu_2}
  -\frac{\varkappa^2}{\nu_1\nu_2}\right) ,
\end{gather*}
where  $\nu_{1,2}=\omega_{1,2}/\eta=  \alpha Z \varepsilon_{1,2} /2$.  The function $t_{12}$  depends on variables $\varepsilon$ and 
$\varkappa$. Then formula \eqref{eq14} reads
\begin{equation}\label{eq17}
\overline{\vphantom{| \cal{A}|^)} |\cal{A}|\,}\!{}^2  = \left(\frac{2\pi\alpha}\eta\right)^4 \frac{2^7 (1+t_{12}^2)
|{\cal{M}}(\varepsilon,\varkappa,t)|^2}{n^3\omega_1\omega_2p(1-e^{-2\pi\xi})}.
\end{equation}
Substituting \eqref{eq17} into \eqref{eq15} and passing on to dimensionless quantities, we obtain the triple differential cross section
\begin{equation}\label{eq18}
\frac{d^3\sigma^{+*}_{ns}}{d\varepsilon d\varkappa dt} = \frac{48 \sigma_{\mathrm{T}}}{Z^2n^3\nu^2_1}
\frac{\varkappa (1+t^2_{12})}{1-e^{-2\pi\xi}}\left| {\cal{M}}(\varepsilon,\varkappa,t)\right|^2 .
\end{equation}
The ionized electron can have energy $\varepsilon$ within the range from $0$ to  $\varepsilon_{\max}=\varepsilon_1-\varepsilon_0$, and the transferred momentum $\varkappa$   is limited by the values $\varkappa_{\min}=\nu_1-\nu_2$ and  $\varkappa_{\max}=\nu_1+\nu_2$.

\begin{figure}[tb]
\begin{minipage}[t]{0.49\textwidth}
\centering\includegraphics[width=0.9\textwidth,angle=0,clip]{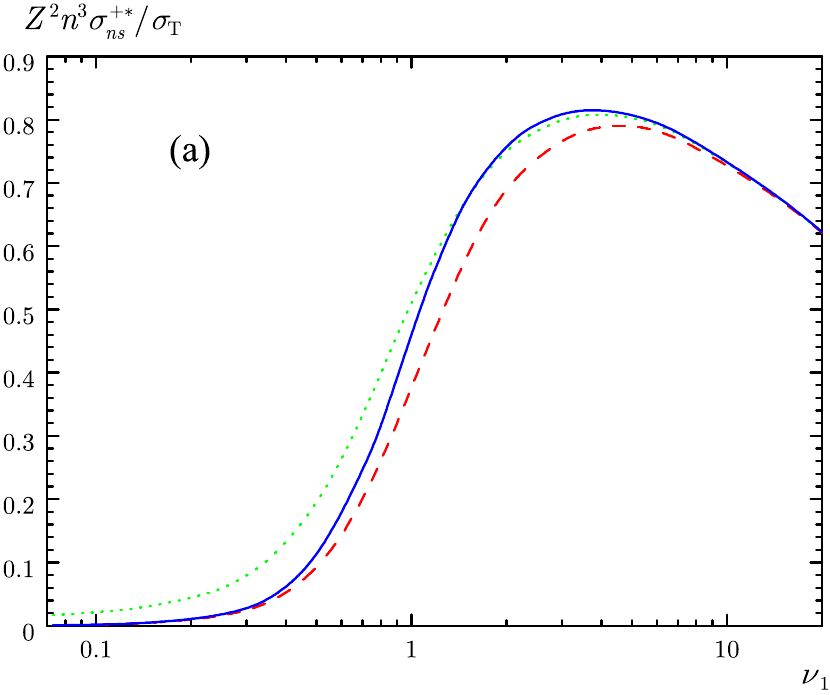}
\end{minipage}
\begin{minipage}[t]{0.49\textwidth}
\centering\includegraphics[width=0.9\textwidth,angle=0,clip]{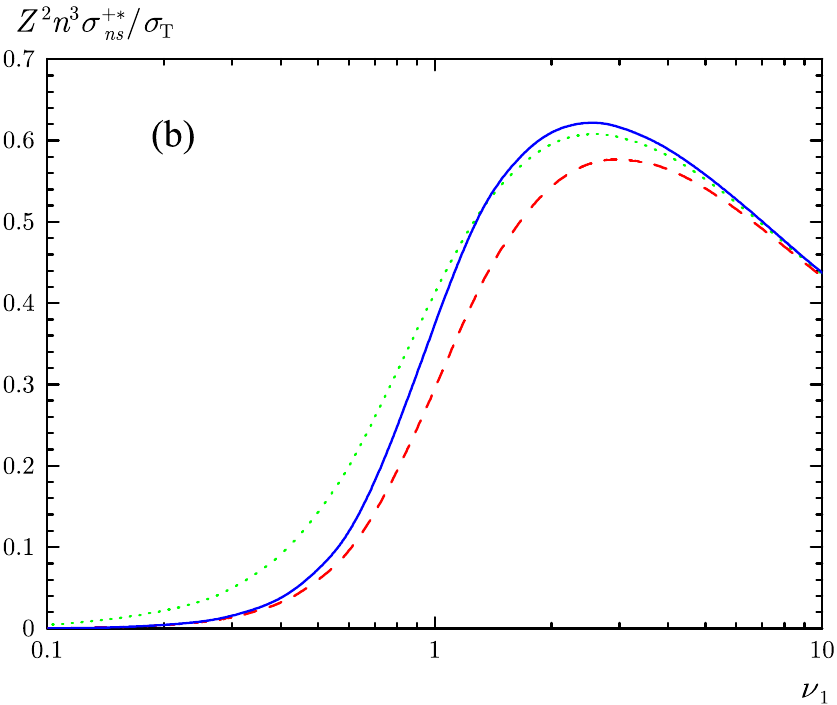}
\end{minipage}
\caption{\label{fig4} Total cross sections \eqref{eq24}  for $n=2$: the contribution of the diagram in Fig.~\ref{fig2}a (dotted curve), the total contribution of the diagrams in Fig.~\ref{fig2}a and  \ref{fig2}b  (dashed curve), the contribution of all diagrams (solid).  (a) $Z=2$,  (b) $Z=10$. }
\end{figure}

\begin{figure}[tb]
\begin{minipage}[t]{0.49\textwidth}
\centering\includegraphics[width=0.9\textwidth,angle=0,clip]{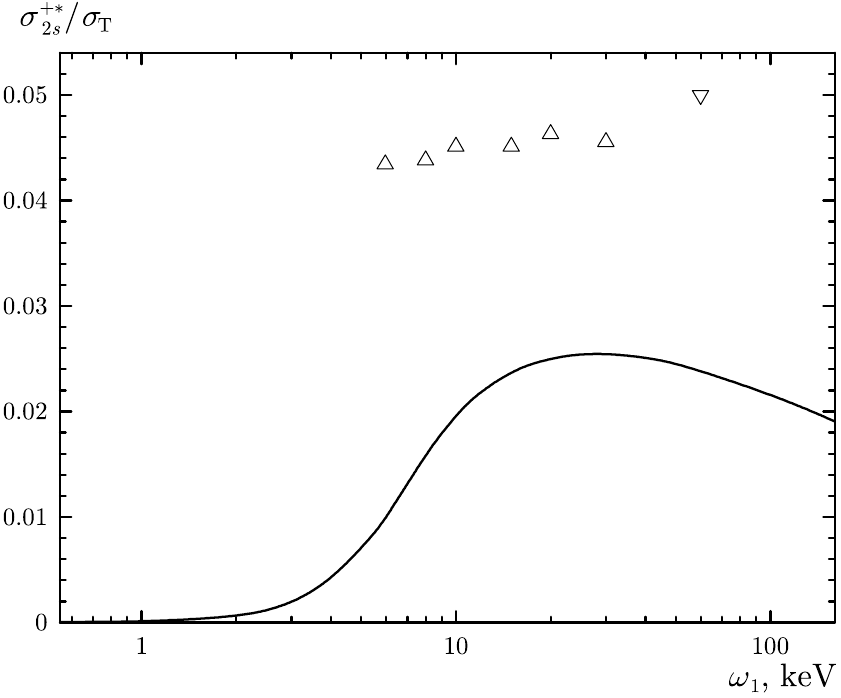}
\end{minipage}
\caption{\label{fig5} Comparison of our calculation (solid curve) with other calculations for helium:  $\vartriangle$  
\cite{15},  $\triangledown$ \cite{16}.}
\end{figure}

Integrating in  \eqref{eq18}  with respect to variables  $t$ and $\varkappa$, we find the energy distributions of   ionized electrons 
$d\sigma^{+*}_{ns}/d\varepsilon$. These distributions are shown in Fig.~\ref{fig3} for $n=2$, $Z=2$, and three energy values $\varepsilon_1=50$ (a), 150 (b), and 500 (c), which correspond to the values $\nu_1=0.365$, $1.095$, and $3.649$. It is seen that at  $\omega_1\lesssim \eta$ the differential cross sections are determined by all four Feynman diagrams, whereas at   $\omega_1\gtrsim 4\eta$ the contribution of the diagram in Fig.~\ref{fig2}a is determinative except for a very small part of the spectrum near $\varepsilon = 0$.  All the curves turn out to be localized in a rather narrow range with respect to the total  interval $0\leqslant \varepsilon \leqslant  \varepsilon_{\max}$ allowed by the energy-conservation law. From Fig.~\ref{fig3} one can also obtain information about the energy distribution of scattered photons $d\sigma^{+*}_{2s}/d\varepsilon_2$, whose curves are symmetric to the plots  $d\sigma^{+*}_{2s}/d\varepsilon$  relative to the vertical line passing through the point $\varepsilon=\varepsilon_{\max}/2$.

Having integrated  \eqref{eq18} with respect to the regions of change of all three variables, we represent the total cross section in the form:
\begin{equation}
\sigma^{+*}_{ns} =\frac{48 \sigma_{\mathrm{T}}}{Z^2n^3\nu^2_1} \int\limits^{\varepsilon_{\max}}_0
\frac{d\varepsilon}{1-e^{-2\pi\xi}} \int\limits^{\varkappa_{\max}}_{\varkappa_{\min}} \varkappa
(1+t^2_{12})d\varkappa  \int\limits^{+1}_{-1} |{\cal{M}}(\varepsilon,\varkappa,t)|^2 dt .  \label{eq24}
\end{equation}
Since the relative contributions of the Feynman diagrams change at characteristic values of  $\omega_1 \sim \eta$, we use the dimensionless energy scale calibrated by the momentum $\eta$. In such units, the photon energies within the range $I \ll \omega_1\ll m$ correspond to the range $\alpha Z/2 \ll \nu_1 \ll (\alpha Z)^{-1}$. The dependence of  cross section \eqref{eq24} on the energy of  incident photon is depicted in Fig.~\ref{fig4} for 
$Z=2$  (a) and $10$ (b).  The cross section multiplied by $Z^2$  weakly depends on $Z$. As in Fig.~\ref {fig3}, it is seen here that for $\nu_1 \lesssim 1$ all Feynman graphs should be taken into account, whereas in the range  $\nu_1\gtrsim 4$ it is enough to consider the graph in  Fig.~\ref{fig2}\emph{a} only. The behavior of  $\sigma^{+*}_{2s}$  qualitatively repeats the predictions of  Schnaidt's formula \eqref{eq1}: the cross section is suppressed  at $\nu_1 \ll 1$, grows rapidly in the transition range $\nu_1 \sim 1$,  and falls at high energies $\nu_1 \gg 1$.

In Fig.~\ref{fig5}, it is given a comparison of our total cross sections $\sigma^{+*}_{2s}$  for  helium atom with similar cross sections calculated in works~\cite {15,16} for transition into the entire $L$-shell in the energy range  $6~\mbox{keV}\leqslant \omega_1 \leqslant 60$~keV. Since in the range of asymptotically high energies $(\omega_1\gg\eta)$ the cross section for ionization  with transition into the $2p$-state is an order of magnitude smaller than the cross section for ionization  with transition into the $2s$-state (see~\cite{12,13, 14}), the cross sections  from works \cite{15,16} should be reduced by approximately 10\% when compared with our $\sigma^{+*}_{2s}$. Characteristically, the calculations of work \cite{15} are practically independent of the photon energy $\omega_1$,  while  work \cite{16} predicts even increase of the cross section at higher energies.

\begin{figure}[t]
\begin{minipage}[t]{0.49\textwidth}
\centering\includegraphics[width=0.9\textwidth,angle=0,clip]{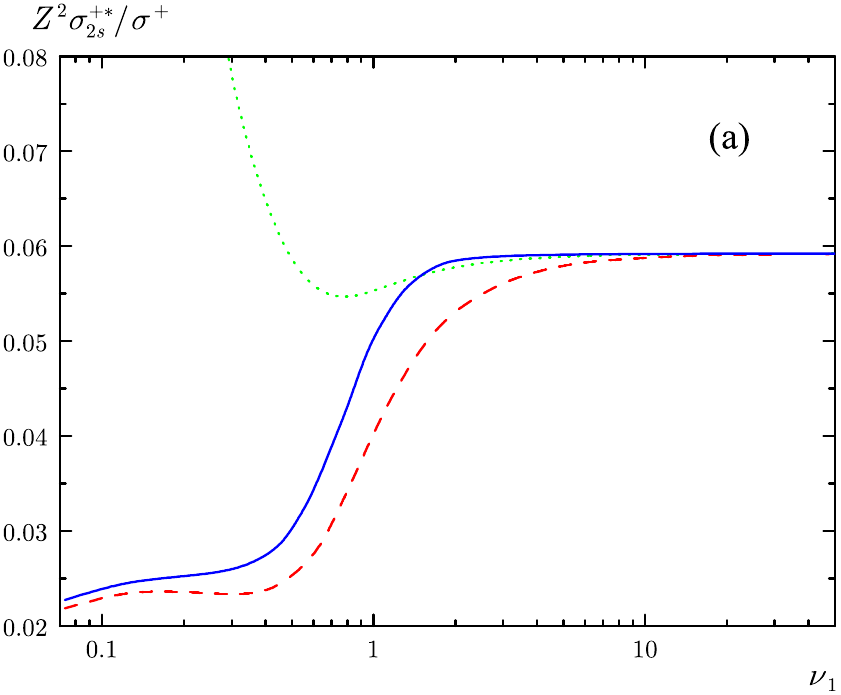}
\end{minipage}
\begin{minipage}[t]{0.49\textwidth}
\centering\includegraphics[width=0.9\textwidth,angle=0,clip]{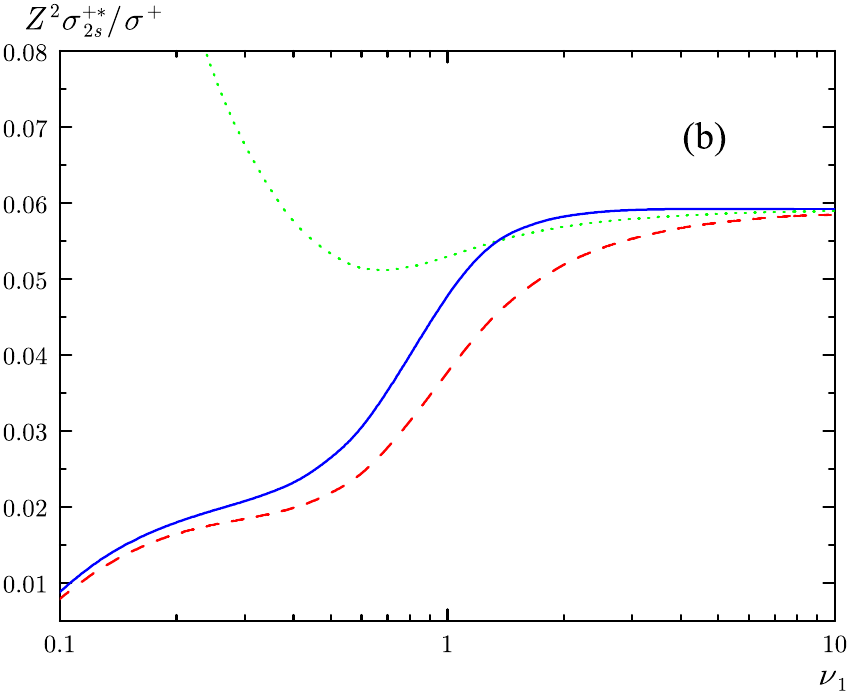}
\end{minipage}
\caption{\label{fig6} 
Ratio of cross sections $Z^2R_{2s}$: the contribution of the diagram in Fig.~\ref {fig2}a
(dotted curve), the contribution of the diagrams in Fig.~\ref{fig2}a  and  \ref{fig2}b  (dashed), and the contribution of all diagrams (solid). 
(a) $Z=2$,  (b)  $Z=10$.}
\end{figure}

The ratio $R_{nl}= \sigma^{+*}_{nl}/\sigma^{+}$  of  the cross sections  for ionization with excitation and ordinary ionization in Compton scattering   from helium-like ions in a wide range of photon energies is of experimental interest. Within the framework of our consideration $\sigma^+ = 2\sigma^+_{1s}$, where $\sigma^+_{1s}$ is described by Schnaidt's formula \eqref{eq1}. In Fig.~\ref{fig6},  it is  shown the behavior of   quantity $Z^2R_{2s}$ in the domain $\alpha Z/2 \ll \nu_1 \ll (\alpha Z)^{-1}$  for $Z = 2$ (a) and $10$ (b). In the range of small  $\nu_1 \ll 1$, the cross-section ratio is small and exhibits a strong dependence on $Z$. In the transition region $\nu_1 \sim 1$, the quantity $Z^2 R_{2s}$ depends weakly on $Z$ and grows rapidly, reaching its maximum value almost at $\nu_1 \gtrsim 2$. Earlier, this universal limit for all $Z$ was obtained for asymptotically high energies  $\nu_1 \gg 1$ \cite{13,14}. Although the cross sections for  both ordinary ionization and ionization with excitation were calculated using the rough Born approximation for  ionized electron \cite{13,14}, the corresponding cross-section ratio \eqref{eq3} has the range of  applicability much broader than could initially be assumed. This is due to the interference of  contributions from the interelectron interaction in the final state of  atom and the exchange interaction, and also due to the rapid decrease of these contributions with increasing photon energy.

\section{Conclusion}

In this paper, the process of Compton scattering from helium-like ions with simultaneous excitation of the $ns$-state of the residual ion is considered. The calculation of differential and total cross sections is performed for photons with energy $I\ll\omega_1\ll m$. In this energy range, it is sufficient enough to use the  ${\bf A}^2$-approximation for  electron-photon interaction and the nonrelativistic approximation for wave functions. The interelectron interaction is taken into account within the framework of perturbation theory with respect to the small parameter $1/Z$. The numerical calculations showed that in the energy range $\omega_1\lesssim \eta$ it is necessary to take into account the electron-electron interaction both in the initial and final states of   atom (the diagrams in Fig.~\ref{fig2}a, \ref{fig2}b, and the exchange diagrams). In the energy range $\omega_1\gtrsim \eta$, the contribution of the diagram in Fig.~\ref{fig2}a, which describes the electron-electron interaction in the initial state, dominates. In the same energy range, the quantity $Z^2\sigma^{+*}_{2s}/\sigma^+$  is the universal function of  $\nu_1=\omega_1/\eta$ for all $Z$ such that  $\alpha Z \ll 1$.

\end{document}